\documentclass[unsortedaddress,preprint,aps,show pacs,showkeys]{revtex4}
\usepackage{graphicx}
\usepackage{dcolumn}
\usepackage{amsmath}
\usepackage{amssymb}
\usepackage{amsfonts}
\usepackage{bm}
\usepackage{color}
\newcommand{\be}{\begin{equation}}
\newcommand{\ee}{\end{equation}}
\newcommand{\ba}{\begin{eqnarray}}
\newcommand{\ea}{\end{eqnarray}}
\begin{document}
\title{A maximum-entropy approach to\\
the adiabatic freezing of a supercooled liquid}
\author{Santi Prestipino$\footnote{Electronic mail: {\tt sprestipino@unime.it}}$}
\affiliation{Universit\`a degli Studi di Messina, Dipartimento di Fisica e di Scienze della Terra, Contrada Papardo, I-98166 Messina, Italy\\and\\CNR-IPCF, Viale F. Stagno d'Alcontres 37, I-98158 Messina, Italy}
\date{\today}
\begin{abstract}
I employ the van der Waals theory of Baus and coworkers to analyze the fast, adiabatic decay of a supercooled liquid in a closed vessel with which the solidification process usually starts. By imposing a further constraint on either the system volume or pressure, I use the maximum-entropy method to quantify the fraction of liquid that is transformed into solid as a function of undercooling and of the amount of a foreign gas that could possibly be also present in the test tube. Upon looking at the implications of thermal and mechanical insulation for the energy cost of forming a solid droplet within the liquid, I identify one situation where the onset of solidification inevitably occurs near the wall in contact with the bath.
\end{abstract}
\maketitle

\section{Introduction}
\setcounter{equation}{0}
\renewcommand{\theequation}{1.\arabic{equation}}

Liquid freezing is a widely studied phenomenon, especially under equilibrium conditions where the temperature $T$ and the pressure $P$ of the system are kept fixed through the contact with a bath. Considerable attention has also been (and still currently is) devoted to the kinetic aspects of freezing, which in a moderately supercooled liquid is initiated by the spontaneous nucleation and growth of a sizable crystal droplet. The usual setting where the initial stages of freezing are studied is again isothermal-isobaric, which assumes a prompt release of the latent heat of solidification to the system environment. In the present study, I keep distinct the two steps by which the freezing process actually develops for a supercooled liquid~\cite{Glicksman,Hindmarsh}, namely a rapid return to the solid-liquid equilibrium temperature (this stage occurs so quickly that there would be no time for a significant transfer of heat to the bath), followed by the slower, diathermal solidification process governed by heat conduction to the bath, which, in effective terms, begins in a later moment. While the late freezing stage has been the focus of many studies (inspiring a whole branch of mathematical physics which goes under the name of ``Stefan problem'', see e.g. \cite{Tabakova}), the initial adiabatic-freezing process has received less attention in the literature, being confined to applied-research areas like Atmospheric Science (where it is studied in connection with the physics of ice accretion and chemicals uptake in hydrometeors~\cite{Macklin,Pruppacher,Stuart,Kostinski}) and Metallurgy (bearing here the name of ``recalescence'', see e.g. \cite{Siegel,Yang}).

Clearly, the initial adiabatic transformation to a two-phase state can only be observed if the process of solid growth takes a time $t_{\rm eq}$ much smaller than the time $t_{\rm dia}$ that would be needed to transfer (essentially by conduction) an energy equal to the latent heat of complete freezing to the surroundings -- differently, solidification would occur without this intermediate two-phase state. This can be represented as
\begin{equation}
t_{\rm eq}\approx V^{1/3}/G\ll t_{\rm dia}\,,
\label{1-1}
\end{equation}
where $V$ is the system volume, $G$ is the long-term growth rate of the solid inside the liquid (measured e.g. in cm/s and likely proportional to the liquid supersaturation for small undercoolings), and $t_{\rm dia}$ is the diathermal-freezing time (see Eq.\,(2) of Ref.\,\cite{Stuart} for an estimate). Whether Eq.\,(\ref{1-1}) is satisfied or not in a concrete case, it would not be easy to say (see, however, below concerning the consequences of thermal insulation for the energetics of droplet formation and the kinetics of crystallization).

At the end of the adiabatic stage, a {\em microsegregated} solid-liquid mixture with definite proportions of the two phases is formed, which later undergoes complete crystallization through the exchange of energy and possibly volume with the colder bath~\cite{Glicksman2,Glicksman3,Voorhees,Marsh}. A fine-grained, uniform distribution of the solid within the liquid is a frequent outcome of recalescence (see e.g. the discussion in \cite{Feuillebois}), as confirmed experimentally for water by Hindmarsh et al.~\cite{Hindmarsh} and well known to metallurgists. When a molten alloy (or a multicomponent liquid whatsoever) is cooled, the growing solid phase usually forms a porous matrix through which the residual liquid can flow. The reactive medium made of the solid matrix and the residual liquid is called a mushy zone. This type of semisolid, solid-liquid mixture occurs for many freezing conditions forming dendrites, see e.g. \cite{Worster}. The full conversion of the mushy zone into a compact solid occurs through a coarsening mechanism (also called Ostwald ripening), involving mass and heat diffusion, and requires some time during which the temperature of the system stays constant at the equilibrium freezing temperature.

From a thermodynamic point of view, the driving force of the return of the metastable liquid to equilibrium is entropy maximization, insofar as the system and (when present) the volume reservoir are treated as parts of a larger thermally isolated system. For this system, the final equilibrium state is the one with the maximum entropy under the given internal constraints~\cite{Callen}. In the mushy zone, however, entropy is not yet at a maximum because of the large amount of interfacial energy trapped in the interstices of the solid matrix. Hence, the state of the system immediately after recalescence is not the one prescribed by thermodynamics but rather a compromise dictated by kinetic considerations. While not being completely realistic, a study of adiabatic freezing solely based on the maximum-entropy principle will anyway help to analyze general trends of thermodynamic variables with supercooling. This would be especially true for rare-gas fluids, having a weakly anisotropic solid-liquid interface tension, and in the low-supercooling regime where, due to a low nucleation rate, the size attained by solid grains before impinging other grains will be larger.

In maximizing entropy, besides the adiabaticity of the system boundaries, additional constraints to be accounted for may regard the total system mass (which is here assumed to be conserved) and its volume {\em or} pressure, depending on the physical context. To make progress, I shall represent the system characteristics by a specific {\em model} of simple-fluid thermodynamics, {\em viz.} the phenomenological theory of Baus and coworkers~\cite{Daanoun,Coussaert}, which, though being of little quantitative value, at least provides a reasonable close-form entropy function for each phase. With this tool at hand, I shall demonstrate that the equilibrium state eventually attained after completion of the adiabatic process is indeed inhomogeneous, with both solid and liquid present in calculable proportions.

This paper is organized as follows. After introducing the model and the method in Section II, I analyze three different physical situations of adiabatic freezing in Section III, for each deriving a number of numerical results. Then, in Section IV these results are exploited to see whether the energy cost associated with the formation of the solid-liquid interface could be a stumbling block to crystallization under isolated conditions. Some concluding remarks are presented in Section V.

\section{Model and Method}
\setcounter{equation}{0}
\renewcommand{\theequation}{2.\arabic{equation}}

The simplest set-up for the study of adiabatic freezing is the following. Consider a $N$-particle liquid which completely fills a closed rigid container of volume $V$. The liquid, initially in stable equilibrium at the melting/freezing temperature $T_m$, is then gently cooled until the temperature $T_{\rm in}<T_m$ is reached. At this point, the pressure and energy of the liquid ($P_{\rm in}$ and $E_{\rm in}$) can be accessed, at least in principle, from the metastable branches of its mechanical and thermal equations of state. Now imagine to remove the contact with the thermostat and to induce then (e.g. by a mechanical shock) the irreversible decay of the system to equilibrium (this transformation is both adiabatic and isochoric, hence energy-conserving). We want to determine in which equilibrium state the system will eventually settle down. This question may be answered by appealing to the maximum-entropy principle. Envisaging the possibility that part of the system could remain liquid, I denote by $E_l,V_l$, and $N_l$ the energy, volume, and particle number of the liquid fraction. If the fundamental relations (entropy functions) of the liquid and solid phases are known, the total system entropy reads
\begin{equation}
S_{\rm tot}=S_l(E_l,V_l,N_l)+S_s(E_{\rm in}-E_l,V-V_l,N-N_l)\,,
\label{2-1}
\end{equation}
assuming weak coupling between the phases and taking the internal constraints into account. The necessary conditions for the maximum of $S_{\rm tot}$ are three, one for each liquid state variable, and are equivalent to requiring that the temperature, the pressure, and the chemical potential of the solid and liquid components be equal. We then see that (unless $N_l=0$ or $N$ at the point of maximum of (\ref{2-1})) the final equilibrium state lies on the solid-liquid coexistence locus, as empirically observed. In order for Eq.\,(\ref{2-1}) to be really useful, however, the functions $S_l$ and $S_s$ are to be made explicit.

The mean-field theory by Baus and coworkers~\cite{Daanoun,Coussaert} provides a convenient framework for discussing adiabatic freezing at a semiquantitative level. It is meant for a system of particles interacting through a spherically-symmetric potential $u(r)$ given by the hard-sphere potential plus a short-range attractive tail,
\begin{equation}
u(r)=\left\{
\begin{array}{ll}
+\infty & ,\,\,\,r<\sigma \\
-\epsilon\,\phi(r/\sigma) & ,\,\,\,r\ge\sigma
\end{array}
\right.\,,
\label{2-2}
\end{equation}
where $\sigma$ is the particle-core diameter, $\epsilon$ is the depth of the attractive well, and $\phi(x)>0$ gives the shape of the well in terms of the scaled interparticle distance $x=r/\sigma$. In the same spirit of the van der Waals theory, the repulsive and attractive potential terms separately concur to build up the system Helmholtz free energy, which is taken to be
\begin{equation}
F(T,V,N)=F_R(T,V,N)+F_A(T,V,N)\,,
\label{2-3}
\end{equation}
where $F_R(T,V,N)\equiv F_{id}(T,\alpha V,N)$ is the free energy of $N$ fictitious non-interacting particles in a fraction $\alpha$ of the total system volume, and
\begin{equation}
F_A(T,V,N)=\frac{N}{2}\int_V{\rm d}^3r\,\overline{\rho}({\bf r})u_A(r)\,,
\label{2-4}
\end{equation}
where $\overline{\rho}({\bf r})$ is the local number density experienced by a reference particle in the origin and $u_A$ is the attractive potential. The description of the model is complete after specifying $\alpha$ and $\overline{\rho}({\bf r})$ for each phase. For a fluid phase, one assumes
\begin{equation}
\alpha=1-\frac{\rho}{\rho_0}\,\,\,\,{\rm and}\,\,\,\,\overline{\rho}({\bf r})=\frac{N}{V}\equiv\rho\,,
\label{2-5}
\end{equation}
with $\rho_0\sigma^3=1/\sqrt{2}+3/(4\pi)$ being a rough estimate of the maximum density accessible to a disordered system~\cite{Coussaert}. For a solid phase, the choice goes to
\begin{equation}
\alpha=\left[1-\left(\frac{\rho}{\rho_{CP}}\right)^{1/3}\right]^3\,\,\,\,{\rm and}\,\,\,\,\overline{\rho}({\bf r})=\sum_{j>1}\delta^3({\bf r}-{\bf R}_j)\,,
\label{2-6}
\end{equation}
where $\rho_{CP}$ is the number density at close packing ($\rho_{CP}\sigma^3=\sqrt{2}$ for a FCC crystal) while $\{{\bf R}_j\}$ are the lattice sites. Upon making the further approximation of discarding the contributions to $F_A$ from particles beyond the first coordination shell, one arrives at the following expression for the free energy per particle:
\begin{equation}
f(T,\rho)=\left\{
\begin{array}{ll}
k_BT\left[\ln(\rho\Lambda^3)-1\right]-k_BT\ln\left(1-\rho/\rho_0\right)-
2\pi\epsilon\,\rho\sigma^3\int_1^{+\infty}{\rm d}x\,x^2\phi(x) & ,\,\,\,{\rm for\,\,a\,\,fluid} \\
k_BT\left[\ln(\rho\Lambda^3)-1\right]-3k_BT\ln\left[1-\left(\rho/\rho_{CP}\right)^{1/3}\right]-
(z_1/2)\epsilon\,\phi\left[(\rho_{CP}/\rho)^{1/3}\right]  & ,\,\,\,{\rm for\,\,a\,\,solid}
\end{array}
\right.,
\label{2-7}
\end{equation}
where $\Lambda\propto T^{-1/2}$ is the thermal wavelength and $z_1$ is the lattice coordination number. It is now straightforward to derive the entropy functions of the solid and fluid phases from Eq.\,(\ref{2-7}). By eliminating $T$ in favor of $e$, we obtain (up to an overall constant):
\begin{equation}
s(e,v)=\left\{
\begin{array}{ll}
k_B\ln(v-v_0)+(3/2)k_B\ln\left(e+a/v\right) & ,\,\,\,{\rm for\,\,a\,\,fluid} \\
3k_B\ln\left(v^{1/3}-v_{CP}^{1/3}\right)+(3/2)k_B\ln\left\{e+(z_1/2)\epsilon\,\phi\left[\left(v/v_{CP}\right)^{1/3}\right]\right\} & ,\,\,\,{\rm for\,\,a\,\,solid}
\end{array}
\right.
\label{2-8}
\end{equation}
with $a=2\pi\epsilon\sigma^3\int_1^{+\infty}{\rm d}x\,x^2\phi(x)$.

In order to single out at a given $T$ the most stable phase as a function of $\rho$, we should plot the two free energies in Eq.\,(\ref{2-7}) vs. $v=\rho^{-1}$ and then use the common-tangent construction. Alternatively, we may plot $\mu$ as a function of $P$ for fixed $T$ and then look for (i) the crossing between the solid and fluid branches, and (ii) the ``swallowtail'' accompanying any isostructural phase transition if present. A typical outcome of this procedure can be seen in Fig.\,1, showing the phase diagram for $\phi(x)=x^{-6}$ and $z_1=12$ (the solid is a FCC crystal). We see that a phase diagram of the standard simple-fluid type emerges in this case, which is enough for characterizing adiabatic freezing by the maximum-entropy method.

Lastly, I describe the procedure by which the minimum of a convex multivariate function, here minus a total entropy, is computed. First a rough minimization of the objective function is attempted by the simulating-annealing algorithm~\cite{Kirkpatrick}, which generates a random walk in state space which eventually brings to the sought minimum. For a convex function, this method is guaranteed to give the absolute minimum, that is the only minimum present, up to an error which decreases with the number of steps in the walk. Next, assuming that we got close to the extremum, a second optimization cycle is started with the gradient-descent method~\cite{Press}, which eventually leads to the desired target state with high precision.

\section{Results}
\setcounter{equation}{0}
\renewcommand{\theequation}{3.\arabic{equation}}

In the present Section, I report and carefully analyze the properties of the inhomogeneous state attained by a supercooled liquid after its adiabatic relaxation to equilibrium, assuming the theory sketched in Section II. Two different experimental situations are discussed, depending on whether a constraint is put on the system volume or pressure. In the constant-volume case, I consider the further possibility that a fixed amount of a foreign gas is present in the container. These cases are analyzed separately in the following.

\subsection{Constant volume}

Consider first a $N$-particle liquid filling completely a closed rigid vessel of volume $V$. Initially at coexistence conditions, the liquid is subsequently driven metastable by slow cooling and then, after removal of the bath, violently perturbed in order to bring it to equilibrium. The final equilibrium state will maximize the total entropy (\ref{2-1}) (assuming no role for the vapor in the process, which is correct as long as the final pressure is larger than the triple-point value). Now, we specialize to a system described by Eq.\,(\ref{2-8}), with $\phi(x)=x^{-6}$ and $z_1=12$. Initially, the liquid temperature and pressure are $T_m$ and $P_m$, defining a point on the solid-liquid coexistence locus. In this state, the specific volume $v_m\equiv V/N$ can be obtained from
\begin{equation}
\frac{k_BT_m}{v_m-v_0}-\frac{a}{v_m^2}=P_m\,.
\label{3-1}
\end{equation}
When the liquid is brought to $T_{\rm in}<T_m$ at constant volume, its energy changes to
\begin{equation}
e_{\rm in}=\frac{3}{2}k_BT_{\rm in}-\frac{a}{v_m}\,.
\label{3-2}
\end{equation}
With these starting conditions, the state eventually reached by the system after disconnecting the bath and inducing solid nucleation is the one yielding the maximum of
\begin{eqnarray}
\frac{S_{\rm tot}}{Nk_B}&=&n_l\left[\ln(v_l-v_0)+\frac{3}{2}\ln\left(e_l+\frac{a}{v_l}\right)\right]
\nonumber \\
&+&(1-n_l)\left\{3\ln\left[\left(\frac{v_m-n_lv_l}{1-n_l}\right)^{1/3}-v_{CP}^{1/3}\right]+\frac{3}{2}\ln\left[\frac{e_{\rm in}-n_le_l}{1-n_l}+6\epsilon\left(\frac{v_{CP}(1-n_l)}{v_m-n_lv_l}\right)^2\right]\right\}\,,
\nonumber \\
\label{3-3}
\end{eqnarray}
where $e_l=E_l/N,v_l=V_l/N$, and $n_l=N_l/N$. The outcome of the maximization procedure are the thermodynamic variables characterizing the liquid fraction of the system in the final state. Any $0<n_l<1$ testifies of a {\em partial crystallization} of the liquid, hence of the stable coexistence of solid and liquid at equilibrium. Indeed, one easily derives from (\ref{2-1}) that the necessary conditions for the maximum of $S_{\rm tot}$ are the equality of $T,P,\mu$ between the phases, as it may be checked {\it a posteriori} from the values of the temperature, the pressure, and the chemical potential of the solid and liquid fractions in the computed equilibrium state.

I studied in detail the case $T_m=0.8\,T_c$. In Fig.\,2 the final values ($T_{\rm fin}$ and $P_{\rm fin}$) of the system temperature and pressure are reported as a function of $T_{\rm in}$. Both quantities are smaller than the respective initial-state values, $T_m$ and $P_m$; however, they too provide coordinates of points on the solid-liquid coexistence line. We also see that $T_{\rm fin}>T_{\rm in}$, i.e., the liquid heats up during the transformation, and the energy needed to the purpose clearly comes from the latent heat of freezing released during solidification. The solid fraction grows practically linearly with $T_m-T_{\rm in}$ (data not shown), up to about 20\% for $T_{\rm in}=0.1\,T_m$ (note that there is no lower limit to undercooling in the present theory). As expected, the entropy increase in the transformation is larger the smaller $T_{\rm in}$.

\subsection{Constant volume with a foreign gas in the vessel}

Let us now suppose that the liquid is prepared at $T_m$ and $P_m$ by exposure to a gaseous atmosphere (e.g. air), and that a small amount of gas gets trapped in the rigid vessel when sealing it. We then have a liquid in equilibrium with an immiscible gas in a container of fixed volume. For simplicity, I describe the gas as ideal and monoatomic, composed of $N_g=x_gN$ particles. The initial specific volume of the liquid, $v_m$, is still given by Eq.\,(\ref{3-1}) but we now have $v_m<V/N\equiv v_{\rm tot}$. Specifically,
\begin{equation}
v_{\rm tot}=v_m+\frac{x_gk_BT_m}{P_m}\,.
\label{3-4}
\end{equation}
As before, we imagine that the liquid and the gas are cooled very slowly until $T_{\rm in}$ is reached. At this point, the volume $v_{\rm in}$ of the liquid is determined by minimizing the total Helmholtz free energy. This leads to the equation
\begin{equation}
\frac{k_BT_{\rm in}}{v_{\rm in}-v_0}-\frac{a}{v_{\rm in}^2}=\frac{x_gk_BT_{\rm in}}{v_{\rm tot}-v_{\rm in}}\,,
\label{3-5}
\end{equation}
which just represents the equality of pressures between the liquid and the gas. After removing the bath, we induce solid nucleation by a mechanical shock, and wait for the system to reach equilibrium. The state eventually attained is such as to maximize the total entropy
\begin{eqnarray}
\frac{S_{\rm tot}}{Nk_B}&=& n_l\left[\ln(v_l-v_0)+\frac{3}{2}\ln\left(e_l+\frac{a}{v_l}\right)\right]+(1-n_l)\left\{3\ln\left(v_s^{1/3}-v_{CP}^{1/3}\right)+\frac{3}{2}\ln\left[e_s+6\epsilon\left(\frac{v_{CP}}{v_s}\right)^2\right]\right\}
\nonumber \\
&+& x_g\left\{\ln\left[\frac{v_{\rm tot}-n_lv_l-(1-n_l)v_s}{x_g}\right]+\frac{3}{2}\ln\left[\frac{e_{\rm in}-n_le_l-(1-n_l)e_s}{x_g}\right]\right\}
\label{3-6}
\end{eqnarray}
with
\begin{equation}
e_{\rm in}=\frac{3}{2}(1+x_g)k_BT_{\rm in}-\frac{a}{v_{\rm in}}\,.
\label{3-7}
\end{equation}
Upon maximizing $S_{\rm tot}$, one obtains the values of $e_l,v_l,n_l,e_s,v_s$ which provide a complete description of the equilibrium state. The five conditions for the maximum of (\ref{3-6}) are equivalent to requiring the same temperature and pressure for the liquid, the solid, and the foreign gas in the final state (namely, $T_l=T_s=T_g\equiv T_{\rm fin}$ and $P_l=P_s=P_g\equiv P_{\rm fin}$); furthermore, also the chemical potentials of the liquid and solid fractions should be the same (i.e., $\mu_l=\mu_s$), indicating that the adiabatic decay of the metastable state eventually results in a stable coexistence between solid and liquid.

In order to characterize adiabatic freezing, useful quantities to be monitored as a function of $T_{\rm in}$ are: the temperature and pressure of the solid-liquid mixture at equilibrium, $T_{\rm fin}$ and $P_{\rm fin}$; the volume of the mixture, $v_{\rm mix}=n_lv_l+(1-n_l)v_s$, as compared to $v_{\rm in}$; and the entropy of the mixture, in comparison with the entropy of the supercooled liquid. For $T_m=0.8\,T_c$, I examined a number of $x_g$ values in the range from $0.0001$ to 1. For example, in Fig.\,3 the values of $T_{\rm fin}$ and $P_{\rm fin}$ are plotted for $x_g=0.001$ and 0.1. Compared to the case where no gas is present, we see little differences for small to moderate undercoolings. However, below $\widetilde{T}\simeq 0.45\,T_m$ and as far as $x_g\ll 1$, we see a sharp change of slope in all curves, which is related to an abrupt crossover in the gas pressure at $T_{\rm in}$ (i.e., the right-hand side of Eq.\,(\ref{3-5})), from large to very small values. The crossover temperature $\widetilde{T}$ is roughly obtained by putting the left-hand side of Eq.\,(\ref{3-5}) to zero for $v_{\rm in}=v_{\rm tot}$ (i.e., only below $\widetilde{T}$, the gas volume at $T_{\rm in}$ is a significant portion of the total volume). It would be interesting to see whether a similar behavior is observed in a real liquid in the deeply supercooled regime. The pairs $(T_{\rm fin},P_{\rm fin})$ for various $T_{\rm in}$ values and for $x_g=0.001$ were reported on the phase diagram in Fig.\,1, so as to confirm that the final equilibrium states are indeed coexistence states. In the top panel of Fig.\,4, the solid fraction $n_s=1-n_l$ is plotted for $x_g=0.001$ and 0.1. It steadily increases with $T_m-T_{\rm in}$, at an almost constant rate only provided $x_g$ is not too small. In the panel below, the entropies of the mixture and the supercooled liquid are reported. For the same two $x_g$ values, Fig.\,5 shows the final volume of the solid-liquid mixture, in comparison with the volume of the supercooled liquid. A clear crossing between the curves is found for a certain $x_g$-dependent value $T_\times$ of $T_{\rm in}$. While above $T_\times$ the decay of the metastable state is accompanied with the system contraction, the opposite (i.e., an expansion) occurs below $T_\times$.

\subsection{Constant pressure}

As a third example, I consider a supercooled liquid which relaxes to equilibrium under conditions that are simultaneously isobaric and adiabatic. This can be realized by conceiving a non-rigid and adiabatic boundary between the system and an environment characterized by the same pressure in all states (i.e., a volume reservoir).

The energy function of the reservoir is clearly $E_r(S_r,V_r,N_r)=-PV_r+f(S_r,N_r)$, where $P$ is a constant and $f$ is an unknown function. This is tantamount to say that the entropy function is of the form
\begin{equation}
S_r(E_r,V_r,N_r)=g(E_r+PV_r,N_r)\,,
\label{3-8}
\end{equation}
for a convenient function $g$. When a system with entropy $S(E,V,N)$ is in contact with a volume reservoir, the equilibrium state of the composite system is such as to maximize the total entropy
\begin{equation}
S_{\rm tot}=S(E,V,N)+g(E_{\rm tot}-E+P(V_{\rm tot}-V),N_r)\,.
\label{3-9}
\end{equation}
It is easy to check from the latter equation that a necessary condition for equilibrium is that the system pressure be also $P$. If, moreover, the system boundary is adiabatic, the only way the system can exchange energy with the reservoir is pressure work, that is $\Delta E=-P\Delta V$. We now ask what is the total-entropy variation $\Delta S_{\rm tot}$ resulting from the transition of the system of interest from an initial state, $(E_{\rm in},V_{\rm in},N)$, to a final state, $(E_{\rm fin},V_{\rm fin},N)$. Considering that $E_{\rm fin}+PV_{\rm fin}=E_{\rm in}+PV_{\rm in}$, we end up with
\begin{equation}
\Delta S_{\rm tot}=S(E_{\rm fin},V_{\rm fin},N)-S(E_{\rm in},V_{\rm in},N)\equiv\Delta S\,,
\label{3-10}
\end{equation}
meaning that the final equilibrium state would also maximize the entropy increase of the system alone. The only residual variable in (\ref{3-10}) is e.g. $V_{\rm fin}$, while $E_{\rm fin}=E_{\rm in}-P(V_{\rm fin}-V_{\rm in})$.

For a two-phase system in contact with a volume reservoir, the total entropy reads
\begin{equation}
S_{\rm tot}=S_l(E_l,V_l,N_l)+S_s(E_s,V_s,N-N_l)+g(E_{\rm tot}-E_l-E_s+P(V_{\rm tot}-V_l-V_s),N_r)\,,
\label{3-11}
\end{equation}
prescribing the same pressure $P$ for both phases at equilibrium. Assuming an initial state where only the liquid phase is present, and using the first law of thermodynamics to prove that $E_{l,\rm fin}+E_{s,\rm fin}+P(V_{l,\rm fin}+V_{s,\rm fin})=E_{l,\rm in}+PV_{l,\rm in}$, the total-entropy increase is again reduced to the system-entropy increase, in turn given by
\begin{eqnarray}
\Delta S&=&S_l(E_{l,\rm fin},V_{l,\rm fin},N_l)+S_s(E_{l,\rm in}+PV_{l,\rm in}-E_{l,\rm fin}-PV_{l,\rm fin}-PV_{s,\rm fin},V_{s,\rm fin},N-N_l)
\nonumber \\
&-&S_l(E_{l,\rm in},V_{l,\rm in},N)\,.
\label{3-12}
\end{eqnarray}
If we now write the necessary conditions for the maximum of (\ref{3-12}), which eventually yield the unknowns $E_{l,\rm fin},V_{l,\rm fin},N_l,V_{s,\rm fin}$, we find that they prescribe the same temperature and chemical potential for each phase (i.e., $T_l=T_s$ and $\mu_l=\mu_s$), as well as the equality of both pressures with $P$ ($P_l=P_s=P$). In explicit terms, the function to be maximized is
\begin{eqnarray}
\frac{S_{\rm tot}}{Nk_B}&=&n_l\left[\ln(v_l-v_0)+\frac{3}{2}\ln\left(e_l+\frac{a}{v_l}\right)\right]
\nonumber \\
&+&(1-n_l)\left\{3\ln\left(v_s^{1/3}-v_{CP}^{1/3}\right)+\frac{3}{2}\ln\left[\frac{e_{\rm in}+Pv_{\rm in}-n_l(e_l+Pv_l)}{1-n_l}-Pv_s+6\epsilon\left(\frac{v_{CP}}{v_s}\right)^2\right]\right\}\,,
\nonumber \\
\label{3-13}
\end{eqnarray}
where $v_{\rm in}$ and $e_{\rm in}$ are the specific volume and energy of the supercooled liquid at $T_{\rm in}$,
\begin{equation}
\frac{k_BT_{\rm in}}{v_{\rm in}-v_0}-\frac{a}{v_{\rm in}^2}=P_m\,\,\,\,{\rm and}\,\,\,\,e_{\rm in}=\frac{3}{2}k_BT_{\rm in}-\frac{a}{v_{\rm in}}\,.
\label{3-14}
\end{equation}

As before, the choice was made that $T_m=0.8\,T_c$. At variance with the previous case where $V$ was a fixed constant, an isobaric-adiabatic relaxation of the supercooled liquid to equilibrium now brings the system invariably to the original phase-diagram point, i.e., $T_{\rm fin}=T_m$ and $P_{\rm fin}=P_m$, but in the modified form of a solid-liquid mixture, whose specific volume $v_{\rm mix}$ is different from $v_m$ and always larger than $v_{\rm in}$ (Fig.\,6, upper panel). Once more, the solid fraction in the mixture is found to increase, to all practical purposes, linearly in $T_m-T_{\rm in}$ (Fig.\,6, lower panel). I finally observe that the above results were successfully checked against the method for isenthalpic freezing described in \cite{Glicksman}.

\section{Discussion}
\setcounter{equation}{0}
\renewcommand{\theequation}{4.\arabic{equation}}

Using a mean-field theory for illustrative purposes, I have shown that, under adiabatic conditions, a supercooled liquid transforms into a stable mixture of solid and liquid just for entropic reasons. However, until now thermal insulation was a mere hypothesis, and the question remains as to what conditions should be met in order that the decay of the metastable state can be treated as adiabatic {\em also in the presence of the bath}. Clearly, a two-stage freezing scenario could only be viable provided the alleged adiabatic step is guaranteed to conclude very quickly after the appearance of the first solid nucleus. Hence, there is no way a thorough analysis of adiabatic freezing can get around genuinely kinetic issues (rate of nucleation, growth velocity, etc.), which however lie outside the scope of a pure equilibrium theory (and of the present study as well).

Possible hindrances to effective adiabaticity are of at least two kinds, one system-specific and another of a more general type. The inability to grow the solid phase rapidly would be typical of good glass formers, i.e., systems with sawtooth-like potential-surface topographies. If such a system is undercooled down to a temperature $T_{\rm in}$ which is only slightly above the glass-transition temperature, we expect that the adiabatic-freezing stage of solidification will be skipped altogether and conventional diathermal freezing (directed from the surface inward) will occur instead. A different and more basic form of kinetic bottleneck to adiabatic crystallization will be described below, after including in the description also the energy cost of the interface between the phases.

For a two-phase equilibrium system which is both thermally and mechanically isolated from the environment, the total entropy can be written in the setting -- originally devised by Gibbs himself~\cite{Gibbs} -- where thermodynamic properties are attached also to the dividing surface $\sigma$ between the phases (see e.g. \cite{Rowlinson,Debenedetti}). The energy of a planar interface of area $A$ can generally be written as $E_\sigma=T S_\sigma+\gamma A+\mu N_\sigma$, where $T$ and $\mu$ are those of the coexisting phases and $\gamma$ is the interface free energy (surface tension). By the Gibbs adsorption equation,
\begin{equation}
S_\sigma{\rm d}T+A{\rm d}\gamma+N_\sigma{\rm d}\mu=0\,,
\label{4-1}
\end{equation}
$E_\sigma$ is reduced to just $\gamma A$ when the surface tension is independent of $T$ and $\mu$. Now switching to a spherical inclusion or droplet of the $\beta$ phase in the metastable mother $\alpha$ phase, I make the further approximation that $\gamma$ is radius-independent, indeed a fair assumption only sufficiently close to coexistence (see e.g. \cite{Prestipino1}); moreover, I shall neglect surface-tension anisotropy, which is a small effect anyway for many crystals~\cite{Prestipino2}. With these simplifications, the entropy of the $\alpha+\beta$ system becomes equal to:
\begin{equation}
S_{\rm tot}(e_\beta, v_\beta,N_\beta;E,V,N)=(N-N_\beta)s_\alpha(e_\alpha,v_\alpha)+N_\beta s_\beta(e_\beta,v_\beta)\,,
\label{4-2}
\end{equation}
where $e_\beta$ and $v_\beta$ are the specific energy and volume of the nucleating phase, and $N_\beta$ is the number of particles in the droplet. In Eq.\,(\ref{4-2}), the energy and volume of the mother phase are given by $E_\alpha=E-E_\beta-E_\sigma$ and $V_\alpha=V-V_\beta$ respectively, or
\begin{equation}
e_\alpha=\frac{E_\alpha}{N_\alpha}=\frac{E-N_\beta e_\beta-(36\pi)^{1/3}\gamma(N_\beta v_\beta)^{2/3}}{N-N_\beta}\,\,\,\,{\rm and}\,\,\,\,v_\alpha=\frac{V_\alpha}{N_\alpha}=\frac{V-N_\beta v_\beta}{N-N_\beta}\,,
\label{4-3}
\end{equation}
$E,V,N$ being the state variables of the composite system. For $\gamma=0$, the absolute maximum of (\ref{4-2})-(\ref{4-3}) clearly coincides with the maximum of (\ref{3-3}) for the given $T_{\rm in}$.

The values of the internal variables $e_\beta, v_\beta,N_\beta$ in a (possibly unstable) equilibrium state are obtained from equating the three partial derivatives of $S_{\rm tot}$ to zero. It is then a simple matter to show that these conditions are equivalent to:
\begin{eqnarray}
T_\beta(e_\beta,v_\beta)&=&T_\alpha(e_\alpha,v_\alpha)\,;
\nonumber \\
P_\beta(e_\beta,v_\beta)&=&P_\alpha(e_\alpha,v_\alpha)+\frac{2\gamma}{r_\beta}\,;
\nonumber \\
\mu_\beta(e_\beta,v_\beta)&=&\mu_\alpha(e_\alpha,v_\alpha)\,,
\label{4-4}
\end{eqnarray}
where $r_\beta=(3N_\beta v_\beta/(4\pi))^{1/3}$ is the droplet radius. Hence, any cluster of the $\beta$ phase which is in equilibrium with the $\alpha$ phase should have the same temperature and chemical potential as $\alpha$, while the two pressures are different and related by the Laplace equation. In particular, Eqs.\,(\ref{4-4}) would hold for the cluster of $\beta$ phase in the inhomogeneous equilibrium state, associated with the absolute maximum of $S_{\rm tot}$.

Now take $\alpha$ to be the liquid ($l$) and $\beta$ the solid ($s$), and assume these phases are described by Baus' theory. As far as the value of $\gamma$ is concerned, anything reasonable is good, for example the orientationally-averaged interfacial free energy of hard spheres~\cite{Davidchack}, $\gamma=0.561\,k_BT_m/\sigma^2$. I first checked that, upon maximizing (\ref{4-2})-(\ref{4-3}) for a number of $T_{\rm in}$ values by the same numerical method as employed before, the conditions (\ref{4-4}) are fulfilled. For small enough supersaturation, however, the maximum of the total entropy is invariably found at $N_s=0$. In order to see what is going on, it is worth looking at the graph of the function $\Delta S=S_{\rm tot}(e_s,v_s,N_s;E,V,N)-Ns_l(E/N,V/N)$, which represents the entropic advantage of the inhomogeneous system over the supercooled liquid. To simplify it further, $\Delta S$ is projected onto the one-dimensional subspace where $e_s$ and $v_s$ are given the same values as in the point of absolute maximum of $S_{\rm tot}$. We are thus left with a function of $N_s$ only, which is reported in Fig.\,7 for two small values of $N$ ($10^3$ and $10^4$) and a few undercooling temperatures. A glance at Fig.\,7 immediately reveals the existence of a sharp $\Delta S$ maximum for a non-zero $N_s$ value, corresponding to a two-phase equilibrium state. However, a satellite maximum also exists at the origin, which is separated from the former one by an entropic ``barrier'' (the valley between the two peaks), and when the supersaturation becomes sufficiently small the absolute maximum of $\Delta S$ jumps to $N_s=0$. Therefore, solid formation is thermally activated (i.e., it necessitates a favorable density fluctuation) and, for any fixed $N$, there is a minimum undercooling threshold (which however is negligible for macroscopic $N$) to overcome in order that solidification may occur. Below this threshold, the assumption of a rapid yet partial solidification of the liquid, which is at the heart of the present calculation, should be rejected -- since no solid component, here modeled for simplicity as consisting of one single block, is found in the equilibrium state -- and the onset of solidification is necessarily at the system surface in contact with the bath. Upon reducing the supersaturation further, the relative maximum for $N_s>0$ disappears and no solid cluster, even only a metastable one, can form. A similar scenario is at work in the canonical-ensemble description of liquid nucleation from vapor~\cite{Calecki}.

Summarizing, the calculations in the present Section were aimed at checking whether the assumption of adiabaticity, which is at the basis of the results of Sections III, can survive the inclusion of the interface-energy contribution in the treatment. A necessary condition for that is a positive maximum of $\Delta S$, which however only appears beyond a certain $N$-dependent undercooling threshold, negligible in the large-size limit. This implies that small-sized liquids must be cooled sufficiently deep in order that freezing may start from the system interior; otherwise, homogeneous solid nucleation is obstructed (not simply activated!) and freezing will proceed diathermally from the outset, i.e., directly from the system boundaries.

\section{Conclusions}

Adiabatic freezing is the first lap of conventional freezing. It is observed whenever the energy released during solid nucleation in the very early stages of crystallization does not reach the thermostat but is almost completely spent in the heating up of the system, whose temperature raises quickly until solid-liquid coexistence is established at the equilibrium freezing temperature. Only later will crystallization proceed diathermally. 

I have studied the adiabatic freezing of a supercooled liquid using the van der Waals theory of Refs.\,\cite{Daanoun,Coussaert}, considering three possible experimental arrangements (constant volume, constant volume with an entrapped foreign gas, and constant pressure). I have clarified that, when heat transfer to the external bath is kinetically hindered, the liquid undergoes partial crystallization just for thermodynamic reasons, i.e., as a result of total-entropy maximization. Although the first outcome of recalescence is not usually the state of maximum entropy, due to the formation of a mushy zone which very slowly evolves to stable equilibrium, at least the trends exhibited by various system quantities with supercooling could roughly be predicted by simple thermodynamic arguments.

In the attempt to unearth hidden hypotheses behind the modeling of the early stages of freezing as effectively adiabatic, I was finally led to consider the entropy of a liquid with a solid droplet inside. I have thus documented the existence, for a small-sized liquid system, of a minimum supersaturation to achieve in order that adiabatic freezing may occur under constant-volume conditions. I defer to a future publication the real-life illustration of some of the features of adiabatic freezing that were highlighted in the present study.

\section*{Acknowledgments}

I gratefully acknowledge many enlightening discussions with Paolo V. Giaquinta. I also express my thanks to Franco Aliotta, Rosina C. Ponterio, Franz Saija, and Cirino Vasi (CNR-IPCF, Messina) for introducing me to the fascinating world of adiabatic freezing. I am also grateful to an anonymous Referee who helped me to improve the paper considerably by pointing out the limits of a purely thermodynamic approach to adiabatic freezing.

\newpage
%
%
\begin{figure}
\includegraphics[width=12cm]{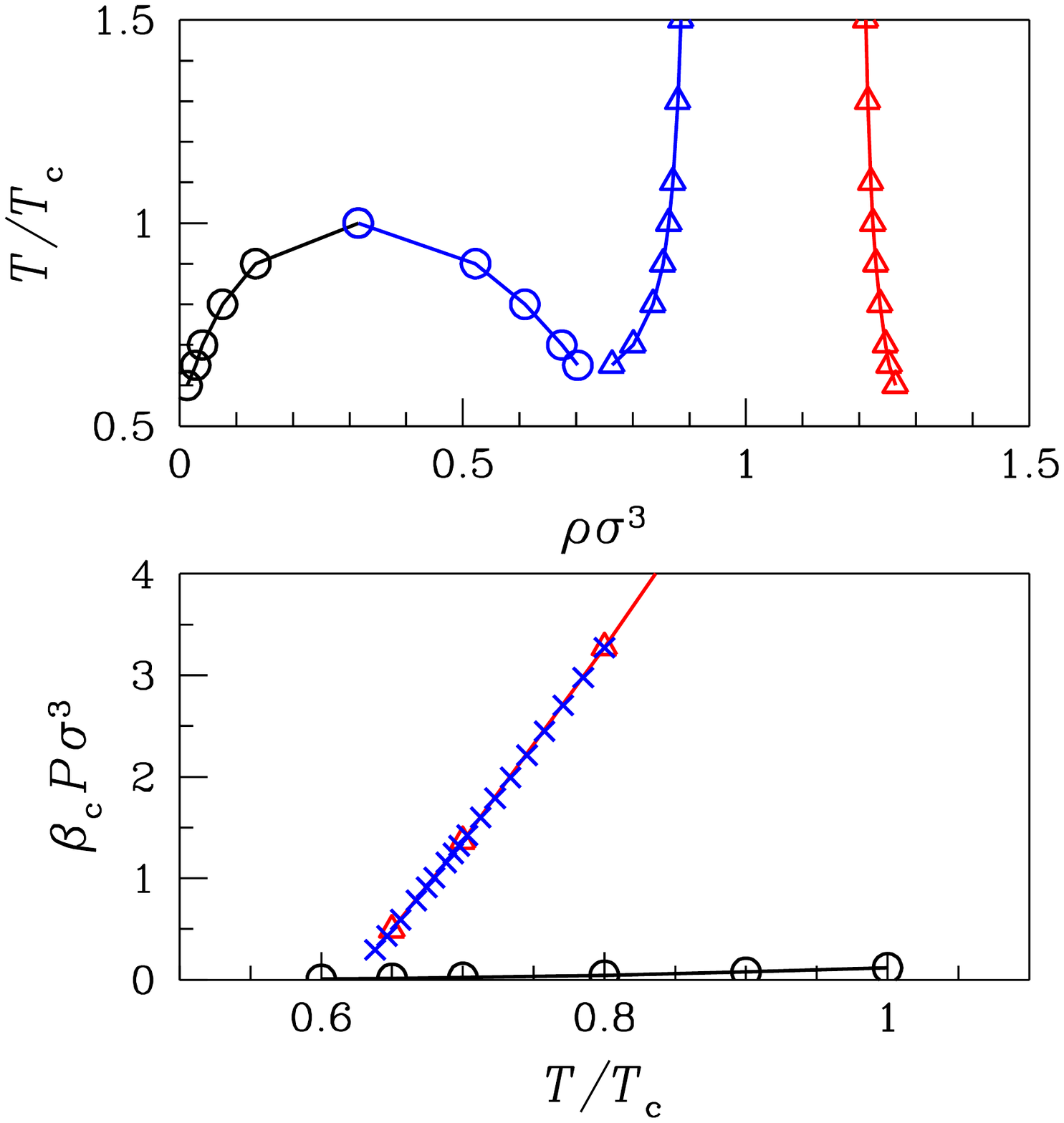}
\caption{(Color online). Theoretical phase diagram for a system of particles interacting through the potential (\ref{2-2}) with $\phi(x)=x^{-6}$ and $z_1=12$. $T_c$ is the critical temperature and $\beta_c=(k_BT_c)^{-1}$. The critical-point coordinates, $\rho_c$ and $T_c$, follow from requiring that the first- and second-order density derivatives of the fluid pressure be simultaneously zero. One thus finds $\rho_c=\rho_0/3$ and $k_BT_c=(8/27)a\rho_0$, with $a=(2\pi/3)\epsilon\sigma^3$. Top: phase diagram on the density-temperature plane, showing the extent of the coexistence regions; the triple temperature is between 0.6 and 0.65 of $T_c$. Bottom: phase diagram on the temperature-pressure plane, reporting as blue crosses also the $(T,P)$ points characterizing the solid-liquid coexistence states borne out of the decay of the metastable-liquid states at various $T_{\rm in}$ values, for $x_g=0.001$ (see Section III.B).}
\label{fig1}
\end{figure}

\newpage
%
%
\begin{figure}
\includegraphics[width=12cm]{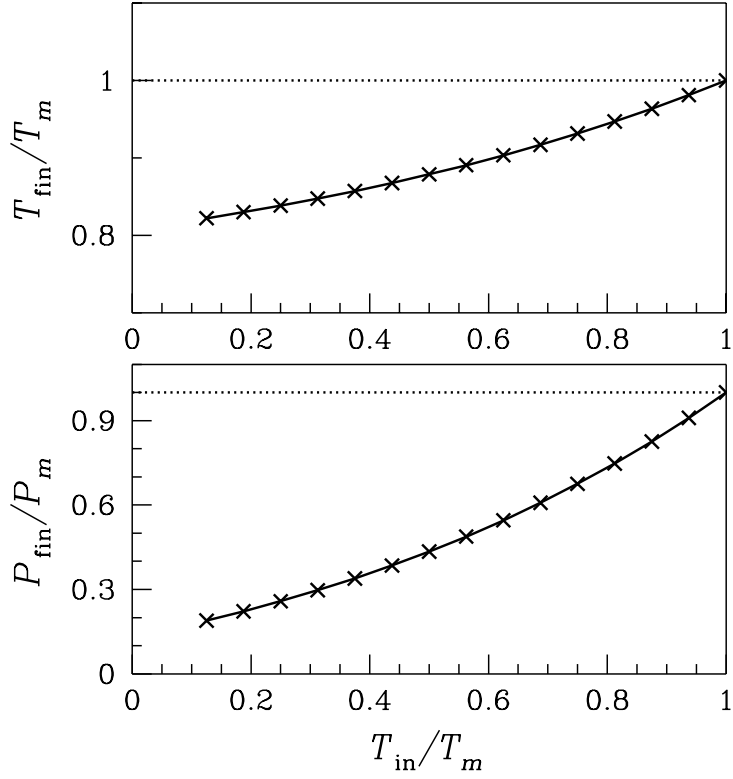}
\caption{Final equilibrium state after the adiabatic decay of the metastable liquid under constant-volume conditions, for $T_m=0.8\,T_c$. Top: temperature; bottom: pressure.}
\label{fig2}
\end{figure}

\newpage
%
%
\begin{figure}
\includegraphics[width=12cm]{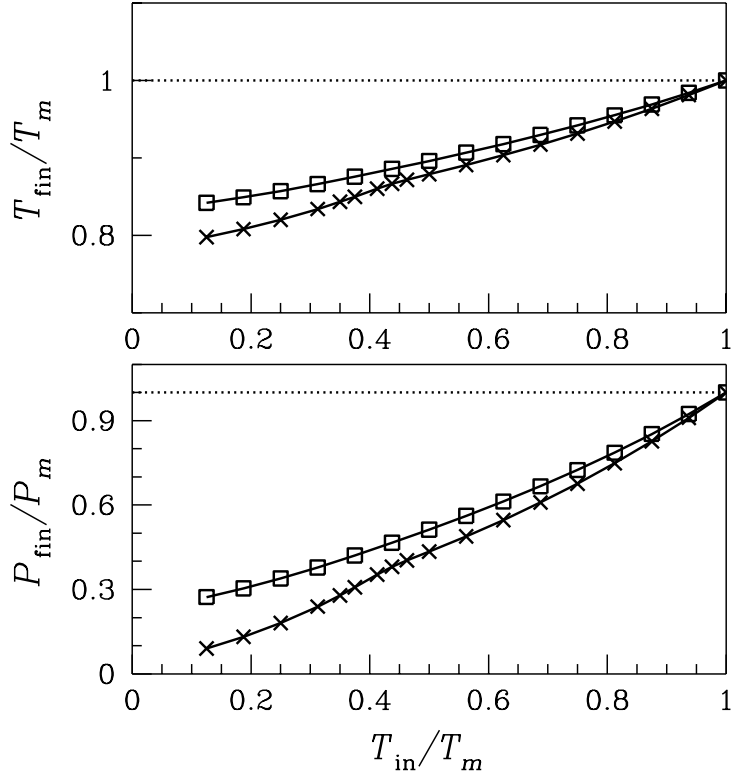}
\caption{Final equilibrium state after the adiabatic decay of the metastable liquid under constant-volume conditions, for $T_m=0.8\,T_c$ and for two different amounts of foreign gas in the vessel (crosses, $x_g=0.001$; squares, $x_g=0.1$). Top: temperature; bottom: pressure.}
\label{fig3}
\end{figure}

\newpage
%
%
\begin{figure}
\includegraphics[width=12cm]{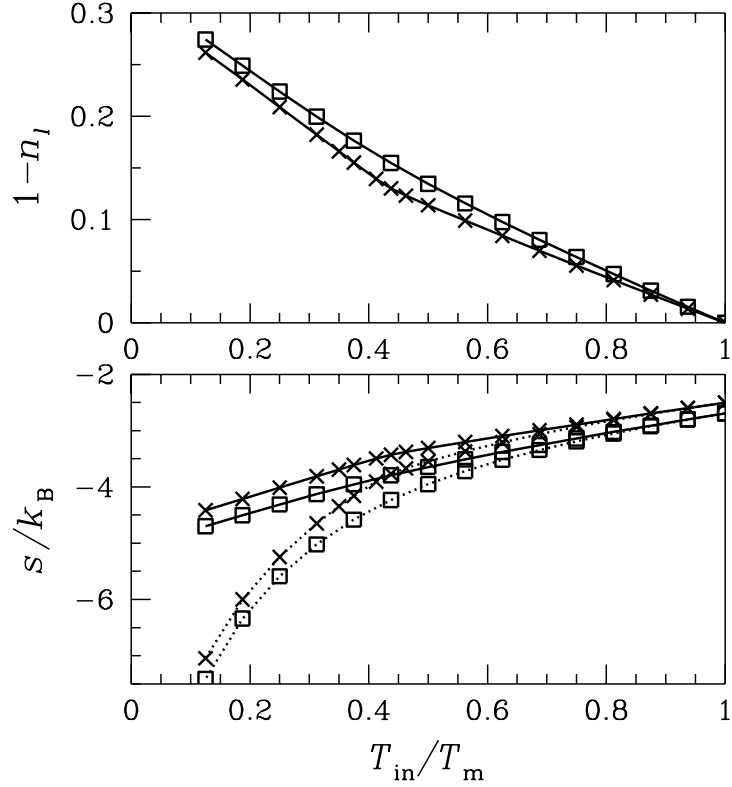}
\caption{Top: Solid fraction in the equilibrium state resulting from the adiabatic decay of the metastable liquid under constant-volume conditions, for $T_m=0.8\,T_c$ and for two different amounts of foreign gas in the vessel (crosses, $x_g=0.001$; squares, $x_g=0.1$). Bottom: Entropy of the solid-liquid mixture at $T_{\rm fin}$ (solid lines) vs. entropy of the supercooled liquid at $T_{\rm in}$ (dotted lines).}
\label{fig4}
\end{figure}

\newpage
%
%
\begin{figure}
\includegraphics[width=12cm]{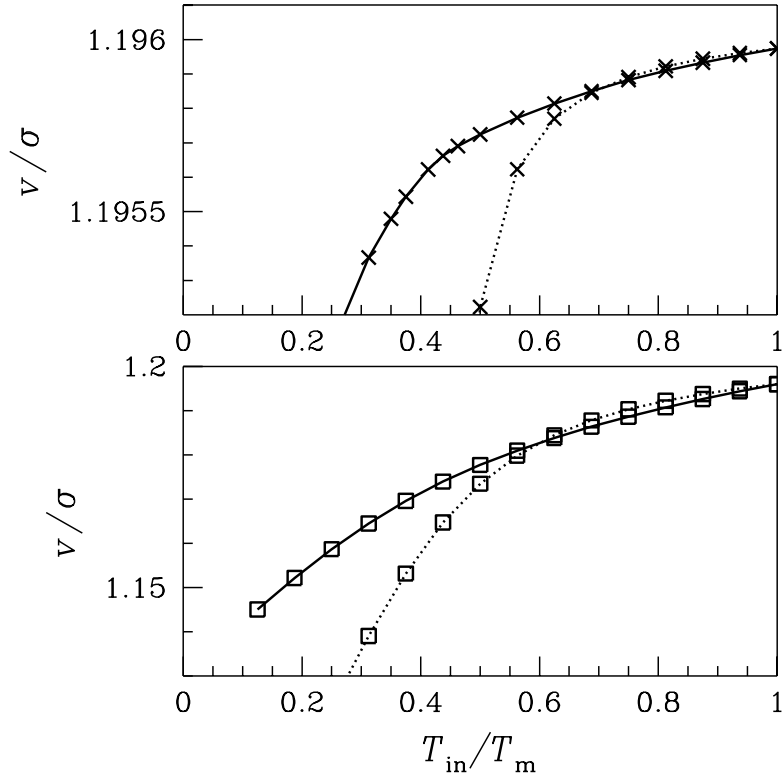}
\caption{Final equilibrium state after the adiabatic decay of the metastable liquid under constant-volume conditions, for $T_m=0.8\,T_c$ and for two different amounts of foreign gas in the vessel (top panel, $x_g=0.001$; bottom panel, $x_g=0.1$). Volume of the solid-liquid mixture (solid lines) vs. volume of the supercooled liquid at $T_{\rm in}$ (dotted lines).}
\label{fig5}
\end{figure}

\newpage
%
%
\begin{figure}
\includegraphics[width=12cm]{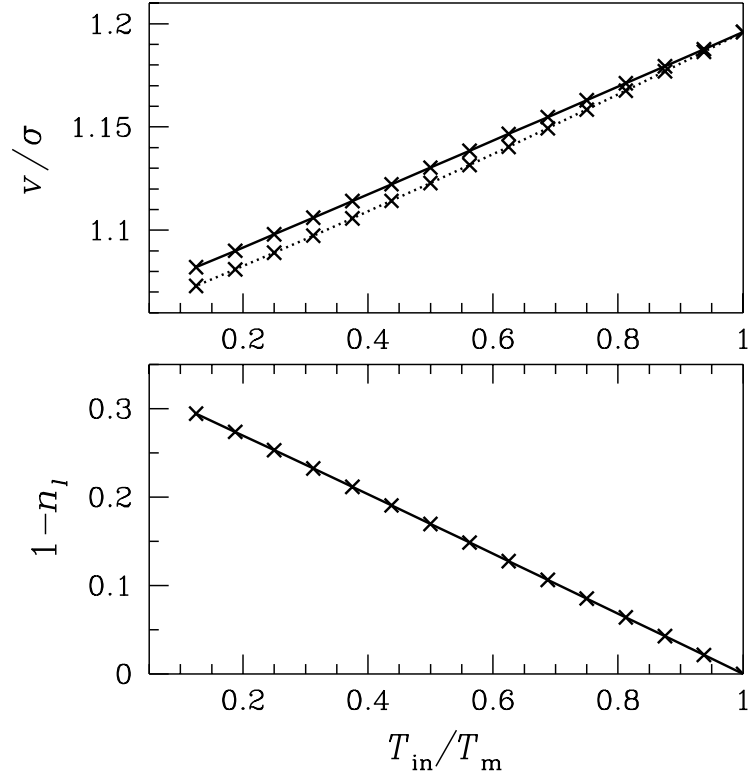}
\caption{Final equilibrium state after the adiabatic decay of the metastable liquid at constant pressure, for $T_m=0.8\,T_c$. Top: volume of the solid-liquid mixture at $T_m$ (solid line) vs. volume of the liquid at $T_{\rm in}$ (dotted line); bottom: solid fraction in the mixture.}
\label{fig6}
\end{figure}

\newpage
%
%
\begin{figure}
\includegraphics[width=12cm]{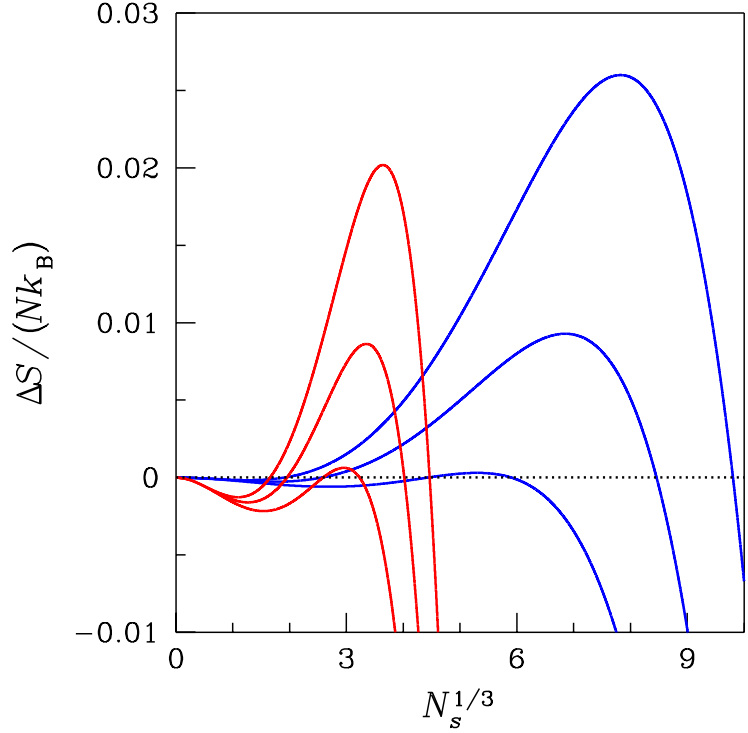}
\caption{(Color online). Difference in specific entropy between the droplet-liquid mixture at $T_{\rm fin}$ and the original metastable liquid at $T_{\rm in}$, as a function of the droplet ``radius'', $N_s^{1/3}$. Two values of $N$ are considered, 1000 (red curves, left) and 10000 (blue curves, right), for $T_m=0.8\,T_c$. For each $N$, various $T_{\rm in}/T_c$ values were considered: from top to bottom, $0.57,0.60,0.63$ for $N=10^3$; and $0.60,0.65,0.70$ for $N=10^4$.}
\label{fig7}
\end{figure}

\end{document}